\documentclass[12pt]{article}
\usepackage{amsmath,amssymb,amsfonts,amsbsy}
\usepackage[english]{babel}
\let\eps\varepsilon

\let\vph\varphi
\let\bb\mathbb
\let\ge\geqslant

\def\p{\partial}

\newcommand{\op}[2]{\vskip0.5ex\par{\bf #1.}{\ \it #2}\vskip0.5ex} 
\begin{document}
\baselineskip=20pt
\begin{center}
{\LARGE\bf Nonlocal symmetries of  integrable two-field divergent evolutionary systems}\footnote{This work is 
supported by the RFBR grant no 05-01-00775. Submitted to Theoretical and Mathematical Physics. }\\[2mm]
A. G. Meshkov\\
Orel State Technical University, Orel, Russia
\end{center}
\begin{abstract}
Nonlocal symmetries for exactly integrable two-field evolutionary systems of the third order have been computed. Differentiation
of the nonlocal symmetries  with respect to spatial variable gives a few nonevolutionary systems for each 
evolutionary system. Zero curvature representations for some new nonevolution systems are presented.
\end{abstract}
\thispagestyle{empty}
\baselineskip6.5mm
\section{Introduction}

This paper is devoted to nonlocal symmetries for the systems obtained in \cite{маг1} via a symmetry classification.
All calculations are simple and do not  require any knowledge of any zero curvature representation or Lax representation.
The investigation gave several new integrable nonevolution systems besides the known Toda lattices.

Kumei's  article \cite{Kum} was probably a pioneering work on generalized symmetries for the sine-Gordon equation
\begin{equation}\label{SG}
v_{tx}=\sin v.
\end{equation}
It was discovered there that one of the symmetries coincides with the modified Korteweg-de Vries equation (mKdV)
\begin{equation}\label{mkdv}
v_t=v_{xxx}+\frac12\,v_x^3
\end{equation}
Rewriting  the sine-Gordon equation in the evolution form $u_t=\p_x^{-1}\sin u$, one can say that this equation is a nonlocal
symmetry of (\ref{mkdv}). Let us consider this problem in detail. 

To simplify all formulas we introduce the function $u=iv_x$, that satisfies the following equation:
\begin{equation}
u_t=u_{xxx}-\frac32\,u^2u_x. \label{mkdv1}
\end{equation}
It is obvious that $u$ is the conserved density for (\ref{mkdv1}) and one can introduce the  nonlocal variable $w=D_x^{-1}u$. 
It can be easily verified that  $e^{w}$ and $e^{-w}$ are the conserved densities  for (\ref{mkdv1}) too. This allows us to introduce
two more nonlocal variables:
$$
w_1=D_x^{-1}e^{w},\ \ w_2=D_x^{-1}e^{-w}. 
$$
Equation (\ref{mkdv1}) possesses the following nonlocal symmetry
\begin{equation}\label{skdv1}
u_\tau= c_1e^{w}+c_2e^{-w}+c_3(w_1e^{-w}+w_2e^{w}),
\end{equation}
where $c_i$ are arbitrary constants (see  \cite{Qiao}, for example). 

If $c_3=0$, then  adopting $w$ in (\ref{skdv1}) as a new unknown function and setting $u=w_x$, we obtain the following well known
integrable equation
\begin{equation}\label{e0}
w_{\tau x}=c_1e^w+c_2e^{-w}.
\end{equation}

Differentiation of equation (\ref{skdv1}) where $c_3=0$, gives $u_{\tau x}=u (c_1e^{w}-c_2e^{-w})$. Excluding here $w$ with the help of the initial
equation, we obtain another integrable equation
\begin{equation}\label{eq0}
u_{\tau x}=u\sqrt{u_{\tau}^2-4c_1c_2}.
\end{equation}
Obviously, the relation  $u=w_x$ connects equations (\ref{e0}) and (\ref{eq0}).

Next, let $c_3\ne0$. Differentiating equation (\ref{skdv1}) and combining the result with the initial equation, one can obtain 
$(u^{-1}u_{\tau x})_x=uu_\tau -2c_3u^{-2}u_x$. Using a dilatation of $\tau$, one can adopt $c_3=-1/2$ and obtain
\begin{align}
u_{\tau xx}=u^{-1}u_x(u_{\tau x}+1)+u^2u_\tau . \label{hyp1a}
\end{align}
If one sets $u_{\tau x}=uz_\tau$, then the hyperbolic system follows:  
\begin{equation}
\label{hyp1}
z_{\tau x}=uu_\tau +u^{-2}u_x, \ \ \ u_{\tau x}=uz_\tau.
\end{equation}

It can be shown that nonlocal equation (\ref{skdv1}) possesses a Lax representation. Hence, all differential consequences of (\ref{skdv1})
have  Lax representations too. So, one integrable evolution equation (\ref{mkdv1}) generates a set of integrable nonevolution equations
(\ref{e0}) -- (\ref{hyp1}).

The canonical approach to obtaining  integrable hierarchies is to fix a Lax operator $L$ and consider various operators  $A$. Usually, 
the  operator $A$ is presented as a polynomial with respect to
positive or negative degrees of the spectral parameter. This gives positive and negative
flows of the Lax equations. There is a remarkable paper on this theme \cite{DS1}, where a general construction of the Lax representations
for the KdV-like equations has been presented in terms of the affine algebras. Later in \cite{D-S}, this construction was described in detail
with proofs and examples provided.

The method that is used here is direct, because it deals with the evolutionary system only. But to prove integrability of a nonlocal symmetry
one must construct the zero curvature representation.
 
\section{Basic notion and notation}
Consider an evolution system with two independent variables $t,x$ and $m$ dependent variables  $u^\alpha$
\begin{equation}\label{EQ}
u_t=K(t,x,u,u_x,\dots,u_n), 
\end{equation}
where  $ K=\{K^\alpha\}$ and $u=\{u^\alpha\},\, \alpha =1,\dots,m$ are infinitely differentiable functions,
$u^\alpha = u_0^\alpha ,\, u_x^\alpha= u_1^\alpha$, $u_k^\alpha=\p^k u^\alpha/\p x^k$. The set of whole  dependent variables  
$u_i^\alpha $ is denoted as $u$ for brevity.
\op{Definition 1}{{\rm (see \cite{Symm},\cite{MSS}).} \sl If the vector function  $\sigma(t,x,u)$ satisfies the equation
\begin{equation}\label{oprU}
(D_t-K_*)\sigma =0,
\end{equation}
where
\begin{align}
&(K_*)^\alpha_\beta =\sum_{k\ge0}\frac{\p K^\alpha }{\p u^\beta_k}D_x^k, \nonumber\\
\label{Diff}
&D_x=\frac{\p}{\p x}+\sum_{\alpha,k\ge0}{}u^\alpha_{k+1} \frac{\p}{\p u^\alpha_k },\ \ \
D_t=\frac{\p}{\p t}+\sum_{\alpha,k\ge0 }^{} \big(D_x^k K^\alpha\big) \frac{\p}{\p u^\alpha_k }, 
\end{align}
then it is said to be the generalized symmetry of system (\ref{EQ}).
}

Here $D_x$ is called the total differentiation operator with respect to $x$, $D_t$ is called the  operator of evolutionary differentiation. 

The order of the differential operator $f_*$ is called the order of the (vector-)function $f$. 

Generalized symmetries are often written as the evolution systems
\begin{equation}\label{evsym}
u_\tau =\sigma (t,x,u),
\end{equation}
where $\tau$ is a new evolution parameter. It is clear that to obtain local integrable equations from (\ref{evsym}) one must find
the symmetries $\sigma$ that do not depend on  $t$ explicitly.

\op{Definition  2}{{\rm (see \cite{Symm},\cite{MSS}).} \sl If for some differentiable functions  $\rho$ and $\theta$  the following equation 
\begin{equation}
D_t\rho (t,x,u)=D_x\theta (t,x,u)   \label{law}
\end{equation}
is satisfied identically for any solution $u$ of system (\ref{EQ}), then relation  (\ref{law}) is called the {\bf local} conservation law of system (\ref{EQ}). 
The function $\rho $ is said to be the conserved density and  $\theta $ is said to be the  density of current. The pair $(\rho ,\theta)$  is said to be the 
 conserved current.}
As the operators $D_t$  $D_x$ are commutative, then the vector $(\rho_0 =D_xf,\,\theta_0 =D_tf)$ with any function $f$ is the  
conserved current for any system. Such currents are called trivial. Conserved currents are always defined by modulo of trivial currents.

Let $(\rho ,\theta)$ be the conserved current, then the following system
\begin{equation}\label{nlo}
w_x=\rho(t,x,u),\ \ \ w_t=\theta(t,x,u) 
\end{equation}
is compatible for any $u$ satisfying equation (\ref{EQ}). The solution of  (\ref{nlo}) is formally written in the form
$w=D_x^{-1}\rho$.
One can consider  $w$ as a new dynamical variable. It is called weakly nonlocal or quasi-local (see \cite{SS}).
We will call such variables the first order nonlocal  variables. Let $(\rho_i, \theta_i)$ be local conserved currents and
 $w_i^{(1)}=D_x^{-1}\rho_i$ be the corresponding  first order nonlocal  variables. If there exist conserved currents depending on
 $w_i^{(1)}$ and, possibly, on local variables, then one can construct new  nonlocal  variables $w_i^{(2)}$ and so on.

The order of  nonlocal  variables is defined inductively. Let the variables $w^{(1)},\dots, w^{(n)}$ be defined till the $n$-th order.
If there exists a {\bf nontrivial}  conserved density $\rho (t,x,u,w^{(1)},\dots, w^{(n)})$ and the $n$-th order variables  $w^{(n)}_i$ can not be removed
by some gauge transformation   $\rho\to \rho +D_x f,\,\theta \to \theta  +D_t f$, then the variable $w=D_x^{-1}\rho (t,x,u,w^{(1)},\dots, w^{(n)})$
is called the $(n+1)$-th order nonlocal  variable.

Operators (\ref{Diff}) are to be prolonged on the  nonlocal  variables $w_i$ in accordance with the following formulas
\begin{equation}\label{prolong}
\hat D_x=D_x+\rho_i \frac{\p}{\p w_i},\quad \hat D_t=D_t+\theta_i \frac{\p}{\p w_i},
\end{equation}
where $(\rho_i,\theta_i)$ are the nonlocal conserved currents corresponding to the nonlocal variables $w_i$.

It is proved  that the operators $\hat D_x$ and $\hat D_t$ are commutative (see \cite{sergeev}, for example).  Hence,
the equation for nonlocal symmetries is  (\ref{oprU}) with the prolonged  operators $\hat D_x$ and $\hat D_t$. 

If prolonged equation  (\ref{oprU}) has a solution depending on  nonlocal  variables, then this solution is called a nonlocal symmetry. 

Differential equations that are interesting for applications have low orders. That is why to obtain interesting nonevolution integrable systems
one ought to consider low order  conserved densities and nonlocal symmetries. We restrict ourselves with considering the local variables
$u_0^\alpha$ and $u_1^\alpha$ only. Moreover, the nonlocal variables are computed till the second order  because the symmetries 
dependent on higher order nonlocal variables are very cumbersome.

If the system takes the following form
\begin{equation}\label{Ediv}
u^\alpha _t=D_xK^\alpha (t,x,u,u_x,\dots,u_{n-1}),
\end{equation}
then it is called divergent. Setting here $u^\alpha=U^\alpha_x$, we obtain the system
\begin{equation}\label{Ept}
U^\alpha _t=K^\alpha (t,x,U_x,\dots,U_{n})
\end{equation}
that is usually called a potential version of system (\ref{Ediv}),  as $U$ is the potential for $u$.

Below we consider systems of the form (\ref{Ept}) with two functions $u$ and $v$. Therefore in the  general formulas
considered above one must change $u^1$ and $u^2$ to $u$ and $v$ respectively.

\section{Nonlocal symmetries} 
Some of the systems found in  \cite{маг1} do not possess any nonlocal symmetries. Other systems possess multi-parametric 
nonlocal symmetries. Arbitrary constants contained in the symmetries are denoted as  $c_i$ or $k_i$. 

{\bf 1.} The system
\begin{equation}
\label{2sys}
u_t=u_3+\frac32\,u_1v_2-\frac34\,{u_1}v_1^2+\frac 1 4\,{u_1}^3, \ \
v_t=-\frac12v_3- \frac34\,(2u_1u_2+u_1^2v_1)+\frac14\,v_1^3
\end{equation}
admits of the following nonlocal symmetry
\begin{equation}\label{nlo1}
\begin{aligned}
u_\tau&= c_2w_2+c_3w_3+c_4(w_4-w_1w_2)+c_5(w_5-w_1w_3)\\
&+c_6(w_3w_4-w_{{2}}w_{{5}})+c_{{7}}(2\,w_{{7}}- w_{{3}}{w_{{1}}}^{2})\\
&+ c_{{8}}(w_{{2}}w_{{7}}+w_{{3}}w_{{6}}+w_{{4}}w_{{5}}-w_{{1}}w_{{3}}w_{{4}}-w_{{1}}w_{{2}}w_{{5}})\\
&+c_{{9}}(w_{{1}} w_{{3}}w_{{4}}+w_{{1}}w_{{2}}w_{{5}}-2\,w_{{3}}w_{{6}}-w_{{4}}w_{{5}}),\\
v_\tau&=c_1w_1-c_2w_2+c_3w_3+c_4w_1w_2-c_5w_1w_3\\
&+c_{{6}}(w_{{3}}w_{{4}}+w_{{2}}w_{{5}}-w_{{8}})
-c_{{7}}(2\,w_{{7}} -2\,w_{{1}}w_{{5}}+w_{{3}}{w_{{1}}}^{2})\\
&+c_{{8}}(w_{{3}}w_{{6}}-w_{{2}}w_{{7}} -w_{{1}}w_{{3}}w_{{4}}+w_{{1}}w_{{2}}w_{{5}})\\
&+c_{{9}}(w_{{3}}w_{{1}}w_{{4}}+w_{{1}}w_{{8}}-2\,w_{{3}}w_{{6}}-w_{{1}}w_{{2}}w_{{5}}),
\end{aligned}
\end{equation}
where
$$
\begin{aligned}
&w_1=D_x^{-1} e^{v},\ \ w_2=D_x^{-1} e^{u-v},\ \ w_3=D_x^{-1} e^{-u-v},\ \ w_4=D_x^{-1}w_2 e^{v},\ \ w_5=D_x^{-1}w_3 e^{v},\\
&w_6=D_x^{-1}w_1w_2 e^v, \ \ w_7=D_x^{-1}w_1w_3e^v,\ \ w_8=D_x^{-1}w_2w_3 e^v. 
\end{aligned}
$$
We have verified that flow (\ref{nlo1}) commutes with flow  (\ref{2sys}) and with the next flow from the same hierarchy:
$$
\begin{aligned}
u_t&=u_5+\frac54 v_4u_1+\frac54 u_3(2v_2-v_1^2)+\frac54v_3(2u_2-u_1v_1)-\frac52 u_2v_2v_1-\frac58 v_2^2u_1\\
&-\frac58u_1v_2(u_1^2+v_1^2)+\frac{1}{32}u_1(5v_1^5-3u_1^4+10u_1^2v_1^2),\\
v_t&=-\frac14 v_5-\frac54u_1u_4-\frac54 u_3(u_2+u_1v_1)-\frac58v_3(u_1^2-v_1^2)+\frac58v_1v_2^2-\frac54u_1u_2v_2\\
&+\frac58u_1u_2(u_1^2+v_1^2)+\frac{1}{32}v_1(5u_1^4-3v_1^4+10u_1^2v_1^2).
\end{aligned}
$$
So, there is a reason to believe that the exact integrability of system (\ref{nlo1}) holds. 

For all systems considered in the paper we have also verified commutativity of the nonlocal flows and of the higher members of the corresponding 
hierarchies. We do not mention it below and we do not write out the  higher members of  hierarchies for brevity.

{\bf 1.a.} Setting in (\ref{nlo1}) $c_i=0,\,i>3$, we obtain the Toda lattice:
\begin{equation}\label{hyp4}
u_{\tau x}= c_2e^{u-v}+c_3e^{-u-v},\ \ \ v_{\tau x}=c_1e^{v}-c_2e^{u-v}+c_3e^{-u-v}.
\end{equation}
In notation of paper \cite{D-S} consider the system
$u_{i,tx} =\exp\left(\sum _j u_j A_{ji}\right)$, where $A_{ji},\ i,j=1,2,3$ is the Cartan matrix of the affine algebra $D^{(2)}_3$ with the following
Dynkin diagram $\circ\!\!<\!\!=\!\!\circ\!\!=\!\!>\!\!\circ$. One  has explicitly:
$$
u_{1,tx} =\exp(2u_1-2u_3),\ \ \ u_{2,tx} =\exp(2u_2-2u_3),\ \ \ u_{3,tx} =\exp(2u_3-u_1-u_2).
$$ 
It is obvious that the functions $p=u_1-u_3,\ q=u_2-u_3$ satisfy the system $p_{tx}=e^{2p}-e^{-p-q},\,q_{tx}=e^{2q}-e^{-p-q}$.
If $c_i\ne0$, then the same system is obtained from (\ref{hyp4}) by the substitution $u-v=2p,\,u+v=-2q$. The constants $c_i$ 
may vanish in (\ref{hyp4}). In particular, if $c_1=0$, then the system decomposes into a pair of the Liouville equations.

System (\ref{hyp4}) can be represented in several forms. For example, choosing $p=w_1$ and $q=w_2$ as the new unknown functions,
we obtain:
$$
p_{\tau x}=p_x(c_1p-c_2q+c_3w), \ \ q_{\tau x}=q_x(2c_2q-c_1p), \ \  \ w_x=p_x^{-2}q_x^{-1}.
$$

{\bf 1.b.} If in (\ref{nlo1}) $c_4=1$ and $c_i=0,i>4$, then there are several possibilities. Consider the following examples.

(1)  Adopting  $p=w_1$ and $q=\ln w_2$ as the new unknown function and setting $c_1=c_3=0$ we obtain:
\begin{equation}\label{eq2}
p_{\tau x}=e^q pp_x,\ \ q_{\tau x}=-e^qpq_x+f(\tau)q_x,
\end{equation}
where $f(\tau)$ is an  integration ``constant''.

(2) Substitution
$$
u=\ln\big(U_x(V_x/U_x)_x\big),\ \  v=\ln U_x,\ \ w_1=U,\ \ w_2=V_x/U_x,\ \ w_4=V
$$
results in another system under condition  $c_3=0$:
\begin{equation}\label{eq3}
U_{\tau x}=c_1UU_x-c_2V_x+UV_x,\ \ V_{\tau x}=c_1VU_x+VV_x +f(\tau)U_x.
\end{equation}
If one considers in this point  $c_3\ne0$, then the result is the third order cumbersome system. 
Notice that if  $c_4=0$ and $c_5\ne0$ in (\ref{nlo1}), then a slightly different substitution results in (\ref{eq3}) again.

(3) Double differentiation of system (\ref{nlo1}) gives  the third order local system 
$$
u_{\tau xx}=(2u_x+z_x)u_{\tau x}+cu_x e^{z}-e^u,\ \ z_{\tau xx}=-(z_x+u_x)z_{\tau x}+c(u_x+2z_x)e^{z}-e^u
$$
for any  $c_1,c_2,c_3$. Here $z=-u-v,\, c=-2c_3$. 

{\bf 1.c.} Adopting $c_7=-1,\,c_i=0,\,i>3$ in (\ref{nlo1}) one can obtain the following system:
\begin{equation}
\begin{aligned}
&u_{txx}=-u_{tx}(p_x+2u_x)+2c_2u_x e^{-p}+2\sqrt{u_{tx}e^p-c_2-c_3e^{-2u}},\\
&p_{txx}=p_{tx}(u_x+p_x)+2c_2(u_x+2p_x) e^{-p}+2\sqrt{u_{tx}e^p-c_2-c_3e^{-2u}},
\end{aligned}  
\end{equation}
where $p=v-u$.

{\bf 2.} The system
\begin{equation}
\begin{aligned}\label{1sys}
&u_t=u_3-3\,v_3+3\,v_2(v_1-2\,u_1)+3\,u_1v_1^2-2\,u_1^3,\\[2mm]
&v_t=-3\,u_3+4\,v_3-3\,u_2(v_1-2\,u_1)+3\,v_1 u_1^2-2\,v_1^3,
\end{aligned}
\end{equation}
admits of the following nonlocal symmetry
\begin{equation}\label{nlo2}
\begin{aligned}
u_\tau& =c_2w_2+c_3w_3+c_4w_4+c_5(2w_2w_3-w_5) +c_6w_6\\
&+c_7(2\,w_3w_4-w_7)+c_8(2\,w_3 w_6-w_9)+c_9w_2(w_2w_3-w_5)\\
&-c_{10}(\,w_5w_6+w_2w_9-2\,w_4w_7+2\,w_3w_4^{2} -2\,w_2w_3w_6),\\
v_\tau &=c_1w_1+c_2w_2+c_4w_1w_2+c_5w_5+c_6(2w_1w_4-w_6)\\
&+c_7w_{{1}}w_{{5}}+c_8(2\,w_1w_7-w_9)+c_9(2\,w_8-w_2w_5)\\
&+c_{10}(w_5w_6-w_2w_9-2w_1w_4w_5+2\,w_1w_2w_7),
\end{aligned}
\end{equation}
where
$$
\begin{alignedat}{3}
&w_1=D_x^{-1} e^{v},&&\ \ w_2=D_x^{-1} e^{-u-v},&&\ \ w_3=D_x^{-1} e^{2u},\\ 
&w_4=D_x^{-1}w_1 e^{-u-v},&&\ \ w_5=D_x^{-1}w_3 e^{-u-v},&&\ \ w_6=D_x^{-1}w_1^2 e^{-u-v},\\
&w_7=D_x^{-1}w_1w_3 e^{-u-v} ,&&\ \ w_8=D_x^{-1}w_2w_3 e^{-u-v} ,&&\ \ w_9=D_x^{-1}w_1^2w_3 e^{-u-v}.  
\end{alignedat}
$$
Let us present some simple local systems that follow from (\ref{nlo2}).

{\bf 2.a.} Setting $c_i=0$ for $i>3$ in (\ref{nlo2}) we obtain the Toda lattice:
\begin{equation}\label{hyp6}
u_{\tau x}=c_2 e^{-u-v}+c_3 e^{2u}, \ \ \ v_{\tau x}= c_1 e^{v}+c_2 e^{-u-v}.
\end{equation}
Let us write the system $u_{i,tx} =\exp\left(\sum _j u_j A_{ji}\right)$, where $A_{ji}$ is the Cartan matrix for the affine algebra $A^{(2)}_4$
 with the following Dynkin diagram $\circ\!\!=\!\!>\!\!\circ\!\!=\!\!>\!\!\circ$. Substitution
$u=2u_1-u_2,\, v=2u_3-u_2$ results in system (\ref{hyp6}) with $c_1=2,c_2=-1,c_3=1$.
System (\ref{hyp6}) can be rewritten in several different forms. For example, the functions $p=w_1, q=w_2$ satisfy the system:
$$ 
p_{\tau x}=p_x(c_1p+c_2q),\ \ q_{\tau x}=-q_x(c_1p+2c_2q+c_3w), \ \ w_x=(p_xq_x)^{-2}.
$$

{\bf 2.b.} If $c_i=0,\,i>4,\,c_4=1$, then double differentiation of system (\ref{nlo2}) gives  the following local system:
\begin{equation}\label{hyp6a}
\begin{aligned}
&u_{\tau xx} = -u_{\tau x}(u_x+v_x)+c(3u_x+q_x)e^{2u}+e^{-u},\ \ (c=c_3),\\
&v_{\tau xx} =v_xv_{\tau x} -u_{\tau x}(u_x+2v_x)+c(u_x+2v_x)e^{2u}+2e^{-u}.
\end{aligned}
\end{equation}
It is obvious that the order of the second equation can be decreased by the substitution $v_x\to v$.
If, under the previous conditions, one chooses  $p=w_1$ and $q=w_2$ as  new unknown functions, then another system follows:
$$
p_{\tau x}=p_x(c_1p+c_2q+pq),\ \ (\ln q_x)_{\tau x}=p_x(c_1+q)-c_3p_x^{-2}q_x^{-2}.
$$

{\bf 2.c.} If $c_4=0,\,c_5=1,\,c_i=0,\,i>5$, then double differentiation of system (\ref{nlo2}) gives  the following local system:

\begin{equation}\label{hyp6b}
\begin{aligned}
&u_{\tau xx} =2u_xu_{\tau x}-v_{\tau x}(3u_x+v_x)+c (3u_x+v_x)e^{v}+3e^{u-v},\\
&v_{\tau xx} =-v_{\tau x}(u_x+v_x)+c (u_x+2v_x)e^{v}+e^{u-v},\ \ (c=c_1).
\end{aligned}
\end{equation}

{\bf 2.d.} If $c_6=1$ and $c_i=0,i>4$, then the  following local system follows
\begin{equation}\label{eq1}
\begin{aligned}
&u_{\tau xx} = -2u_{\tau x}(u_x+q_x)+2c(2u_x+q_x)e^{2u}+2\sqrt{\rule{0pt}{4mm}u_{\tau x}e^{2q}+be^{-2u}-ce^{2(u+q)}},\\ 
&q_{\tau xx} = q_{\tau x}(u_x+2q_x)+\frac{1}{2}c(2q_x-u_x)e^{2u}+\sqrt{\rule{0pt}{4mm}u_{\tau x}e^{2q}+be^{-2u}-ce^{2(u+q)}},
\end{aligned}
\end{equation}
where $b=c_4^2/4-c_2,\, c=c_3,\, q=(v-u)/2$. Notice that in the case $b=c=0$  the order of the first equation can be decreased by the substitution 
$u_x\to u$.

{\bf 3.}  The next system
\begin{equation}\label{3sys}
u_t=u_3+u_1v_2-u_1v_1^2,\ \ \  v_t=(u_2u_1+u_1^2v_1),
\end{equation}
admits of the following nonlocal symmetry
\begin{equation}\label{nlo3}
\begin{aligned}
u_\tau &=c_1 w_1+c_2w_2+c_4(w_4-w_1w_3)+c_5(w_5-w_1w_4)\\
&+c_{{6}}(w_{{6}}-w_{{2}}w_{{3}})+c_{{7}}(w_1w_6-w_2w_4) +c_{{8}}(w_{{8}}-w_{{2}}w_6),\\
v_\tau &= -c_1w_1+c_2w_2+c_3w_3+c_4(w_4+w_1w_3)+c_5w_1w_4\\
&-c_6(w_6+w_{{2}}w_{{3}}) -c_7(w_1w_6+w_{{2}}w_{{4}})-c_{{8}}w_{{2}}w_{{6}},
\end{aligned}
\end{equation}
where
$$
\begin{aligned}
&w_1=D_x^{-1} e^{u-v},\ \ w_2=D_x^{-1} e^{-u-v},\ \ w_3=D_x^{-1} e^{2v},\ \ w_4=D_x^{-1}w_1 e^{2v},\\  
&w_5=D_x^{-1}w_1^2 e^{2v},\ \ w_6=D_x^{-1}w_2e^{2v},\ \ w_7=D_x^{-1}w_1w_2e^{2v},\ \ w_8=D_x^{-1}w_2^2e^{2v}.
\end{aligned}
$$

{\bf 3.a.} In the case $c_i=0,i>3$ the following Toda lattice is obtained:
\begin{equation}\label{nlo8}
u_{\tau x} = c_1e^{u-v}+c_2e^{-u-v},\ \ \ v_{\tau x}= -c_1e^{u-v}+c_2e^{-u-v}+c_3e^{2v}.
\end{equation}
In the new variables $p=u-v,\,q=-u-v$ this system takes the form $p_{\tau x}=2c_1e^p-c_3e^{-p-q},\ q_{\tau x}=-2c_2e^q-c_3e^{-p-q}$. 
This allows to connect system  (\ref{nlo8})  with the affine algebra $C^{(1)}_2$ having the following Dynkin diagram 
$\circ\!\!\!=\!\!\!>\!\!\circ\!\!<\!\!\!=\!\!\!\circ$.
In the case of $c_3=0$ this system decomposes into a pair of the Liouville equations obviously.

{\bf 3.b.} If $c_4\ne0,c_i=0,i>4$, then in the terms of new variables $p=w_1,q=w_3$ system (\ref{nlo3}) takes the following form:
$$
p_{\tau x} = p_x(2c_1p-c_3q-2c_4pq),\ \ (q_x^{-1}q_{\tau x})_x = 2c_3q_x-2c_1p_x+4c_4pq_x+2c_4qp_x+2c_2p_x^{-1}q_x^{-1}.
$$
If one simply doubly differentiates  system  (\ref{nlo3}), then the result is
$$
u_{\tau xx} = u_{\tau x}(2u_x-p_x)-2c_2u_xe^{-p}-c_4e^p,\ \  p_{\tau xx} = 2p_{\tau x}(p_x-u_x)+2c_2e^{-p}(2u_x-3p_x)+2c_4e^p,
$$
where $p=u+v$. Here the order of the first equation can be decreased by the substitution $u_x\to u$.

{\bf 3.c.} If $c_5\ne0$ and the other constants $c_i=0,i>3$, then a double differentiation of system  (\ref{nlo3}) gives
\begin{equation}\label{eq4}
\begin{aligned}
&u_{\tau xx} = u_{\tau x}(2u_x+q_x)-2c_2u_xe^q +\sqrt{aq_{\tau x}e^{2u}+be^{-2q}+2ac_2e^{2u+q}},\\  
&q_{\tau xx} = -2q_{\tau x}(u_x+q_x)-2c_2(2u_x+3q_x)e^q+2\sqrt{aq_{\tau x}e^{2u}+be^{-2q}+2ac_2e^{2u+q}},
\end{aligned}
\end{equation}
where $a=-c_5,b=-c_3c_5,q=-u-v$.

Notice that systems (\ref{eq1}) and (\ref{eq4}) coincide when $c_2=c_3=c_4=0$. This is surprising, because the symmetries of these systems,
i.e. systems (\ref{1sys}) and (\ref{3sys}), are entirely different. A possible explanation is as follows. The system 
$$
u_{\tau xx} = u_{\tau x}(2u_x+q_x) +e^{u}\sqrt{q_{\tau x}},\ \  \
q_{\tau xx} = -2q_{\tau x}(u_x+q_x)+2e^{u}\sqrt{q_{\tau x}}
$$
is Liouvillean and possesses a double sequence of symmetries that are constructed by different integrals.

{\bf 4.} The next system
\begin{equation}\label{4sys}
u_t=u_3+v_1v_2-\frac12\,u_1^3+\frac12\,u_1v_1^2+c_1v_1,\ \ \ v_t=u_2v_1-\frac12\,u_1^2v_1+\frac12\,v_1^3-c_1u_1+c_2v_1
\end{equation}
contains two essential constants that affect the form and quantity of admissible symmetries. It becomes clear if one takes into account
that system (\ref{4sys}) can be obtained from the Ito system by a differential substitution depending on $c_1$ and $c_2$ (see \cite{маг1}). 
If $c_1=0$, then the differential substitution is essentially simplified, and  the differential substitution vanishes when  $c_1=c_2=0$.

{\bf 4.1.} If $c_1=0,\, c_2=0$, then system (\ref{4sys}) is degenerate. In fact, the substitution $u=\ln U_x,\,v=V_xU_x^{-1}$
gives
$$
U_{{t}}=U_{xxx}-\frac32\,\frac {U_{xx}^2}{U_x}-\frac {V_{x}\,V_{xx}\,U_{xx}}{U_{x}^{2}}
+\frac12\,\frac {V_{x}^{2}U_{xx}^{2}}{U_{x}^{3}}+\frac12\,\frac {V_{xx}^2}{U_{x}},\ \ \ V_t=\frac {V_{x}\,U_t}{U_x}.
$$
Hence, $V=F(U)$ and we have:
$$
U_t = U_{xxx}-\frac32\frac{U_{xx}^2}{U_x}+\frac12(F'')^2 U_x^3.
$$
This equation is exactly integrable iff $F^{IV}=0$ (see \cite{MSS}). Hence,  system (\ref{4sys}) with  $c_1=0$, $c_2=0$ is not integrable
in the general case.

{\bf 4.2.} If $c_1=0,\,c_2\ne0$, then  system (\ref{4sys}) admits of the following nonlocal 4-parametric symmetry
\begin{equation}\label{nlo4}
\begin{aligned}
u_\tau &=k_1w_1+k_2w_2+k_3(w_5+2w_2w_3)+k_4(w_1w_5-4w_1w_2+2w_2w_4),\\
v_\tau &= -2k_2e^{-u}v_x-4k_3e^{-u}v_x(2+w_3)-4k_4e^{-u}v_xw_4,
\end{aligned}
\end{equation}
where
$$
\begin{aligned}
&w_1=D_x^{-1} e^{u},\ \  w_2=D_x^{-1} e^{-u}(v_x^2+c_2),\ \ w_3=D_x^{-1} e^{u}w_2,\\ 
&w_4=D_x^{-1} e^{u}w_2w_1,\ \  w_5=D_x^{-1}( 4v_x^2e^{-u} -e^{u}w_2^2).
\end{aligned}
$$

{\bf 4.2.a.} If $k_i=0,i>2$, then differentiation of system (\ref{nlo4}) gives: 
\begin{equation*}
u_{\tau x} =k_1e^{u}+k_2(v_x^2+c_2)e^{-u}, \ \ \ v_{\tau} =-2k_2v_xe^{-u}.
\end{equation*}

{\bf 4.2.b.} If $k_3\ne0, k_4=0$, then one can obtain $k_3=1/4$ by a dilatation of  $\tau$. In this case differentiation of system (\ref{nlo4})
gives the following system:
$$
\begin{aligned}
&u_{\tau x} =\frac14 e^u\left[\left(\frac{v_\tau}{v_x}\right)_x+\frac{v_\tau u_x}{v_x}\right]^2-\frac{1}{2} v_\tau(v_x+c_2v_x^{-1})+k_1e^u-c_2e^{-u},\\
&\big(\ln v_x\big)_{\tau x} =\left[\frac{v_\tau}{v_x^2}(v_{xx}-u_xv_x)\right]_x-(v_x^2+c_2)e^{-u}.
\end{aligned}
$$

{\bf 4.3.} If $c_1\ne0$ in (\ref{4sys}), then a dilatation of  $v,t$ and $x$ gives $c_1=1$:
\begin{equation}\label{4sysa}
u_t=u_3+v_1v_2-\frac12\,u_1^3+\frac12\,u_1v_1^2+v_1,\ \ \ v_t=u_2v_1-\frac12\,u_1^2v_1+\frac12\,v_1^3-u_1+c_2v_1,\tag{\ref{4sys}$a$}
\end{equation}
There are three cases for three different values of $c_2$.

{\bf 4.3.a.} $c_2=-2\eps ,\, \eps =\pm1$. In this case system (\ref{4sysa}) admits the following nonlocal symmetry:
\begin{equation}\label{nlo5}
\begin{aligned} 
u_\tau& =k_1w_1-k_2w_2+k_3w_3+k_4w_4+k_5w_1w_3+k_6(6w_4w_5-6w_3w_6-3w_1+w_{8}), \\ 
v_\tau &=(k_2-\eps k_1)w_1+\eps k_2 w_2+\eps (k_3+k_5w_1)(2v_xe^{-u-\eps v}-w_3)\\
&+k_4(w_3+2\eps vv_xe^{-u-\eps v}-\eps w_4-2v_xe^{-u-\eps v})\\
&+4k_6e^{-u-\eps v}(6vv_x+v_xv^3+3\eps v_xvw_5-3v_xw_5-6\eps v_xv^2-3\eps v_xw_6)\\
&+3k_6(\eps w_1+2w_3w_5+w_7-4w_4+2\eps w_3w_6 -2\eps w_4w_5)-k_6\eps w_{8}.
\end{aligned}
\end{equation}
Here, 
$$
\begin{aligned}
&w_1=D_x^{-1} e^{u+\eps v},\ \ w_2=D_x^{-1}ve^{u+\eps v},\ \ w_3=D_x^{-1} e^{-u-\eps v}(1-\eps v_x^2),\ \ w_5=D_x^{-1} w_3e^{u+\eps v},\\  
&w_4=D_x^{-1}e^{-u-\eps v}\big(v+v_x^2(1-\eps v)\big),\ \ w_6=D_x^{-1}e^{u+\eps v}(vw_3-w_4),\\  
&w_7=D_x^{-1}\big(2ve^{-u-\eps v}(\eps v+v_x^2(2\eps-v) )-e^{u+\eps v}w_3^2\big),\\
&w_{8}=D_x^{-1}\left(2ve^{-u-\eps v}\big(v^2(\eps -v_x^2)-3v+6v_x^2(\eps v-1)\big)+3e^{u+\eps v}w_3(vw_3-2w_4)\right).
\end{aligned}
$$
If $k_5=k_6=0$, then differentiation of system (\ref{nlo4}) gives the following system: 
$$
\begin{aligned}
u_{\tau x}&=(k_1-k_2v) e^{u+\eps v}+k_3e^{-u-\eps v}(1-\eps v_x^2)+k_4e^{-u-\eps v}\big(v +v_x^2(1- \eps v)\big),\\
 v_{\tau x}& =(k_2-\eps k_1+ \eps k_2v ) e^{u+\eps v}-\eps k_3e^{-u-\eps v}(2u_xv_x-2v_{xx}+\eps v_x^2+1)\\
&+k_4e^{-u-\eps v }\big((2v_{xx}-2u_xv_x-1)(\eps v-1) + v_x^2(2\eps-v) \big).
\end{aligned}
$$
If one set here $k_3=k_4=0$, then the triangle system $(u+\eps v)_{\tau x}= \eps k_2e^{u+\eps v}$ follows.

If $k_5=\eps/2$,  and the other constants $k_i=0$, then the following local system follows
\begin{equation}\label{eq5}
\begin{aligned}
p_{\tau x}& = \eps p\sqrt{\eps -\eps p_xq_x},\\
q_{\tau x} &= 2(q_xp_x-1)p_x^{-1} + pp_{xx}(2-q_xp_x)p_x^{-3}+pp_x^{-1}q_{xx}+\eps q\sqrt{\eps -\eps p_xq_x},
\end{aligned}
\end{equation}
where $p=w_1, q=w_3$. Other combinations of constants in (\ref{nlo5}) give very cumbersome systems. 

{\bf 4.3.b.} If $c_2=-2k,\, |k|<1$, then system (\ref{4sysa}) possesses  the following nonlocal symmetry:
\begin{equation}\label{nlo6}
\begin{aligned}
u_\tau &= k_1w_1-k_2w_2+k_3w_3+k_4w_4+k_5(w_4w_1+w_3w_2)+k_6(-w_4w_2+w_1w_3)\\
&-k_7(-w_7+2c^2w_3w_6-2c^2w_4w_5)+k_8(2c^2w_4w_6+2c^2w_3w_5+w_{8}),\\ 
v_\tau &=k_1(cw_2-kw_1)+k_2(kw_2+cw_1)+k_3\Big(2v_x \sin(cv )e^{-u-kv}+w_4c-kw_3\Big)\\
&+k_4\Big(2v_x \cos(cv )e^{-u-kv}-w_3c-kw_4\Big)+2v_x k_5\Big(w_1\cos(c v)+w_2\sin(c v)\Big)e^{-u-kv}\\
&-k_5(kw_3w_2+cw_3w_1-cw_4w_2+kw_4w_1)+2v_x k_6\Big(w_1\sin(c v)-w_2\cos(c v)\Big)e^{-u-kv}\\
&+k_6(cw_1w_4+cw_2w_3-kw_1w_3+kw_2w_4)\\
&-2v_x k_7\Big(\sin(c v)(1-2c^2+2c^2w_6)-2c\cos(c v)(v+k-cw_5)\Big)e^{-u-kv}\\
&+k_7(2kc^2(w_3w_6-w_4w_5)-2c^3(w_4w_6+w_3w_5)+2cw_4-kw_7-cw_{8})\\
&+2k_8v_x \Big(\cos(c v)(1-2c^2+2c^2w_6)+2c\sin(c v)(cw_5-k+v)\Big)e^{-u-kv}\\
&+k_8\Big(2c^3(w_4w_5-w_3w_6)-2kc^2(w_4w_6+w_3w_5)+cw_7+2cw_3+kw_{8}\Big).
\end{aligned}
\end{equation}
Here  $c=\sqrt{1-k^2}$ and
$$
\begin{aligned}
&w_1=D_x^{-1}\cos(cv)e^{u+kv},\ \   w_3=D_x^{-1}\big(\cos(cv-\alpha )-v_x^2\sin(cv)\big)e^{-u-kv}, \\
&w_2=D_x^{-1} \sin(cv)e^{u+kv},\ \  w_4=D_x^{-1}\big(\sin(\alpha -cv)-v_x^2\cos(cv)\big)e^{-u-kv},\\
&w_5=D_x^{-1}\big(w_4\sin(c v)-w_3\cos(c v)\big)e^{u+kv},\ \ w_6=D_x^{-1}\big(w_4\cos(c v)+w_3\sin(c v)\big)e^{u+kv},\\
&w_7=D_x^{-1}\Big(c^2e^{u+kv}\big(\sin(cv)(w_3^2-w_4^2)+2w_3w_4\cos(cv)\big)+\\
&\qquad+e^{-u-kv}\big(c\cos(cv)(2kv_x^2-2vv_x^2+2kv-1)-\sin(cv)(2c^2v_x^2-v_x^2+2c^2v+k)\big)\Big),\\
&w_{8}=D_x^{-1}\Big(c^2e^{u+kv}\big(\cos(cv)(w_3^2-w_4^2)-2w_4w_3\sin(cv)\big)+\\
&\qquad+e^{-u-kv}\big(\cos(cv)(2c^2v_x^2-v_x^2+2c^2v+k)+c\sin(cv)(2kv_x^2-2vv_x^2+2kv-1)\big)\Big),\\
&k=\sin \alpha,\ \ c=\cos \alpha,\ \ -\frac{\pi}{2}<\alpha <\frac{\pi}{2}.  
\end{aligned}
$$
Simple local equations exist under conditions  $k_i=0,i>4$ only:
\begin{equation}\label{hyp9}
\begin{aligned}
u_{\tau x}& =(k_1\cos(cv)+k_2\sin(cv))e^{u+kv}+k_3v_x^2\sin(cv+\alpha )e^{-u-kv}\\
&-k_4v_x^2\cos(cv+\alpha )e^{-u-kv}-\big(k_3\cos(cv)+k_4\sin(cv)\big)e^{-u-kv},\\
v_{\tau x}& = \big(k_1\sin(cv-\alpha )-k_2\cos(cv-\alpha )\big)e^{u+kv}\\
&+(2v_xu_x-2v_{xx}+1)\big(k_3\sin(cv+\alpha )-k_4\cos(cv+\alpha )\big)e^{-u-kv}\\
&-v_x^2\big(k_3\cos(cv+2\alpha )+k_4\sin(cv+2\alpha )\big)e^{-u-kv}.
\end{aligned}
\end{equation}
If $k_3=k_4=0$, then this system decomposes into two Liouville equations in the terms of variables $p=u+ie^{-i\alpha }v,
q=u-ie^{i\alpha }v$.

{\bf 4.3.c.} If $c_2=-a-a^{-1},\ |a|\ne1$, then system (\ref{4sysa})  possesses  the following nonlocal symmetry:
\begin{equation}\label{nlo7}
\begin{aligned}
u_\tau &= -ak_1w_1+k_2w_2+ak_3w_3+k_4w_4+ak_5w_1w_3+k_6w_2w_4\\
&+k_7(w_7+2w_4w_5(a^2-1)^2)+ak_8\big(2w_3w_6(a^2-1)^2+w_{8}\big), \\
v_\tau &= k_1w_1-ak_2w_2 -k_3(w_3-2av_xe^{-u-v/a})-ak_4(w_4-2v_xe^{-u-av})\\
&+k_5w_1(-w_3+2ae^{-u-v/a}v_x)+k_6w_2a(2v_xe^{-u-av}-w_4)\\
&-ak_7\big(w_7+2w_4(a^2-1)(w_5a^2+2a-w_5)\big)\\
&-4ak_7e^{-u-av}v_x\big(2a^2v(a^2-1)+2a-w_5(a^2-1)^2\big)\\
&+k_8\big(2w_3(a^2-1)(w_6-a^2w_6+2a^2)-w_{8}\big)\\
&+4ak_8e^{-u-v/a}v_x\big(w_6(a^2-1)^2+2av(a^2-1)-2a^4)\big).
\end{aligned}
\end{equation}
Here,
$$
\begin{aligned}
&w_1=D_x^{-1}e^{u+v/a},\ \ w_2=D_x^{-1} e^{u+av},\ \ w_3=D_x^{-1} e^{-u-v/a}(a-v_x^2),\\ 
&w_4=D_x^{-1}e^{-u-av}(1-av_x^2),\ \ w_5=D_x^{-1}w_4e^{u+av},\ \ w_6=D_x^{-1}w_3e^{u+av}\\
& w_7=D_x^{-1}\left(4a^2e^{-u-av}\big(v-a^2v-a+v_x^2(a^3v-av+1)\big)-e^{u+av}w_4^2(a^2-1)^2\right),\\
& w_{8}=D_x^{-1}\left(4ae^{-u-v/a}\big(a^3v-a^2-av+v_x^2(-a^2v+a^3+v)\big)-e^{u+v/a}w_3^2(a^2-1)^2\right).
\end{aligned}
$$

If $k_i=0,i>4$, then the following local system follows:
\begin{equation*}
\begin{aligned}
u_{\tau x}&= -ak_1e^{u+v/a}+k_2e^{u+av}+ak_3e^{-u-v/a}(a-v_x^2)+k_4e^{-u-av}(1-av_x^2),   \\
v_{\tau x}&=k_1e^{u+v/a}-ak_2e^{u+av}-k_3e^{-u-v/a}(2au_xv_x-2av_{xx}+v_x^2+a)\\
&-ak_4e^{-u-av}(2u_xv_x-2v_{xx}+av_x^2+1).
\end{aligned}
\end{equation*}
If $k_3=k_4=0$ then this system decomposes into two Liouville equations in the terms of  variables $p=u+av,
q=u+v/a$.

If $k_5=a$ and the other constants $k_i=0$, then the following local system follows
\begin{equation}\label{eq6}
\begin{aligned}
p_{\tau x} &= pqp_x(a^2-1)+2ap\sqrt{a+p_xq_x},\\
q_{\tau x}& = 4a^2(a+p_xq_x)p_x^{-1}+2a^2pp_x^{-1}q_{xx}-2ap_x^{-1}(qp_x+2pq_xp_{xx})\sqrt{a+p_xq_x}\\
&-2a^2pp_{xx}(p_xq_x+2a)p_x^{-3}+(1-a^2)pqq_x,
\end{aligned}
\end{equation}
where $p=w_1,q=w_3$. Other combinations of the constants in (\ref{nlo7}) give more cumbersome systems. 

Notice that all formulas from points 4.3.b and 4.3.c are connected with each other by the transformation $a=k+ic, a^{-1}=k-ic, c=\sqrt{1-k^2}$.
All formulas from point  4.3.a can be obtained from corresponding formulas of point 4.3.b as the limit $k\to\eps=\pm1, c\to0$.
But these calculations are very cumbersome.
In particular, system (\ref{eq6}) is reduced into  (\ref{eq5}) under the substitution  $a=\eps=\pm1$,  $q\to-\eps q$

All remaining systems found in  \cite{маг1} have no nonlocal symmetries or have trivial  nonlocal symmetries that lead to the 
 Liouville equation.

\section{Zero curvature representations}
We present here the matrices $U$ and $V$ realizing zero curvature representations 
$$
U_\tau -V_x+[U,V]=0
$$
for some of the systems connected with  (\ref{1sys}). Spectral parameter is denoted as  $k$ everywhere.

System (\ref{1sys}) can be obtained from the Drinfeld-Sokolov system \cite{D-S}
\begin{equation}
\begin{aligned}\label{ds}
&m_t= m_3-3n_3-3m_x(4m-9n)+3n_x(8m-15n), \\
&n_t=-3m_3+4 n_3+12m_xn+6n_x(m-4n)
\end{aligned}
\end{equation}
by the following differential substitution:
\begin{equation}\label{sub}
m =u_x^2+\frac{1}{2}v_x^2-u_2-v_2, \ \ \ n = u_x^2-u_2.
\end{equation}
First, we write the matrices  $U_0$ и $V_0$ that form the zero curvature representation for system  (\ref{ds}):
\begin{equation}\label{U0}
\begin{aligned}
U_0&=
\begin{pmatrix}
0& 1 & n-m& 0 & 0\\  0& 0& 0& 1 & 0\\ -1 & 0 & 0 & 0 & m-n \\
0& n & 0 & 0 & k \\ 0 & 0 & 1 & 0 &0 
\end{pmatrix},
V_0&=
\begin{pmatrix}
h_{1,x} & h_2 & f_1 & 0 &-5 k \\ 0 & h_{3,x} & -5k & -2h_3 & 0\\ 
-h_1 & 0 & 0 & 5 & -f_1 \\ 5 k & f_2 & 0 & -h_{3,x} & kh_2\\
0 & 5 & h_1 & 0 & -h_{1,x}
\end{pmatrix}.
\end{aligned}
\end{equation}
Here,
$$
\begin{aligned}
&h_1=7n-4m,\ \ h_2= m-3n,\ \ h_3=4n-3m,\\
&f_1=-4m_2+7n_2+4m^2+7n^2-11mn,\\
&f_2=-3m_2+4n_2-8n^2+6mn.
\end{aligned}
$$
Matrices (\ref{U0}) are embedded in $sl(5,\bb C)$.

Performing substitution (\ref{sub}) in matrices  (\ref{U0}) and excluding $u_2$ and $v_2$ from  $U_0$ by a gauge transformation
$U=S^{-1}(U_0S-S_x),\ V=S^{-1}(V_0S-S_t)$, we obtain the zero curvature representation for system (\ref{1sys}):
\begin{equation}\label{U1}
\begin{aligned}
U&=
\begin{pmatrix}
v_x & 1 & 0 & 0 & 0 \\ 0 & -u_x & 0 & 1 & 0 \\ -1 & 0 & 0 & 0 & 0\\ 0 & 0 & 0 & u_x & k \\
0 & 0 & 1 & 0 & - v_x
\end{pmatrix},
V&=
\begin{pmatrix}
\vph_1 & \vph_2 & 0 & -5v_x & -5k \\ 0 & \vph_3 & -5k & \vph_4 & 5kv_x \\
-\vph_5 & 5h & 0 & 5 & 0\\ 5k & 0 & 5kh & -\vph_3 & k\vph_2\\
0 & 5 & \vph_5 & 0 & -\vph_1
\end{pmatrix}.
\end{aligned}
\end{equation}
Here, 
$$
\begin{aligned}
&\vph_1=4v_3-3u_3+3u_2(2u_x-v_x)+3v_xu_x^2-2v_x^3,\\ 
&\vph_2=2u_2-v_2-2u_x^2-2v_x^2+5u_xv_x,\\
&\vph_3=3v_3-u_3 +3v_2(2u_x-v_x)-3u_xv_x^2+2u_x^3,\ \ h=v_x-u_x,\\
&\vph_4=2u_2-6v_2-2u_x^2+3v_x^2,\ \ \vph_5=4v_2-3u_2+3u_x^2-2v_x^2.
\end{aligned}
$$
The system  $\Psi _x=U\Psi $, where $U$ takes the form (\ref{U1}), can be reduced to the following single equation
$$
(\p_x-u_x)(\p_x+u_x)(\p_x-v_x)\p_x(\p_x-v_x)\Psi_5+k\Psi_5=0.
$$
The spectral problem for this equation is obviously nontrivial.

System (\ref{1sys}) is presented in  \cite{D-S}, but in another form  (see table 5, $A^{(2)}_4$). The zero curvature representations 
for this system and corresponding Toda lattice are contained in the same paper. But it was simpler for us to compute these
zero curvature representations anew. Matrix  $U$ for the Toda lattice (\ref{hyp6}) is shown in (\ref{U1}) and $V$ takes the following form:
$$
V=
\begin{pmatrix}
0 & 0 & -c_1e^v & 0 & 0 \\ -c_2 e^{-u-v} & 0 & 0 & 0 &0 \\ 0 & 0 & 0 & 0 & c_1e^v\\
0 & c_3 e^{2u} & 0 & 0 & 0 \\ 0 & 0 & 0 & -k^{-1}c_2e^{-u-v} & 0
\end{pmatrix}.
$$

We have assumed that systems (\ref{hyp6a}) -- (\ref{eq1}) belong to the same hierarchy as system (\ref{1sys}). If this is true, the matrix $U$ is
common for all mentioned systems. The calculations have confirmed our assumption and we present below only the matrices $V$ for the 
mentioned systems.

For system (\ref{hyp6a}):
\begin{equation}\label{U2}
V=
\begin{pmatrix}
0 & 0 & u_{\tau x}-v_{\tau x}-c e^{2u} & 0 & 0 \\ c e^{2u}-u_{\tau x} & 0 & -e^{-u} & 0 &0 \\ 0 & 0 & 0 & -k^{-1}e^{-u} & v_{\tau x}-u_{\tau x}+c e^{2u}\\
0 & c e^{2u} & 0 & 0 & 0 \\ 0 & 0 & 0 & k^{-1}(c e^{2u}-u_{\tau x}) & 0
\end{pmatrix}.
\end{equation}

For system  (\ref{hyp6b}):
\begin{equation}\label{U3}
V=
\begin{pmatrix}
0 & 0 & -ce^{v} & 0 & 0 \\ ce^{v}-v_{\tau x} & 0 & 0 & 0 &0 \\ 0 & 0 & 0 & 0 & ce^{v}\\
-e^{u-v} & u_{\tau x}-v_{\tau x}+ce^{v} & 0 & 0 & 0 \\ 0 & k^{-1}e^{u-v} & 0 & k^{-1}(ce^{v}-v_{\tau x}) & 0
\end{pmatrix}.
\end{equation}

For system  (\ref{eq1}):
\begin{equation}\label{U4}
V=
\begin{pmatrix}
0 & 0 & -2q_{\tau x}-ce^{2u} & 2k^{-1} e^{2q} & 0 \\ ce^{2u}-u_{\tau x} & 0 & -2r & 0 & 2e^{2q} \\
0 & 0 & 0 & -2k^{-1}r & 2q_{\tau x}+ce^{2u} \\ 0 & ce^{2u} & 0 & 0 & 0 \\ 0 & 0 & 0 & k^{-1}(ce^{2u}-u_{\tau x}) & 0
\end{pmatrix}.
\end{equation}
Here $r=\sqrt{\rule{0pt}{4mm}u_{\tau x}e^{2q}+be^{-2u}-ce^{2(u+q)}}$ and the substitution $v=u+2q$ must be performed in the
matrix $U$ (see (\ref{U1})).

\medskip
\section*{Conclusion}
As it was mentioned above, each nonlocal symmetry presented in this paper is a symmetry for the system under consideration as
well as for its higher analogue. This gives grounds to believe that all presented systems are exactly integrable. But this assumption 
must be proved, of course. Such proofs have been presented for systems (\ref{hyp6a}) -- (\ref{eq1}). For other
systems this problem should be further investigated.


\begin{thebibliography}{99}
\itemsep 0mm
\bibitem{маг1} A. G. Meshkov, {\it Fundamentalnaya i Prikladnaya Matematika,} {\bf 12}:7, (2006), 141--161,\ (in Russian).
\bibitem{Kum} S. Kumei, {\it J. Math. Phys.}, {\bf 16}:12 (1975), 2461--2468.

\bibitem{Qiao} Zhijun Qiao, arXiv:nlin/0201065v1 [nlin.SI], 31.01.2002.
%
\bibitem{DS1} V. G. Drinfeld, V. V. Sokolov, {\it Sov. Math. Dokl.,} {\bf 23} (1981), 457.)
%
\bibitem{D-S}  V. G. Drinfeld, V. V. Sokolov,  {\it Lie algebras and equations of Korteweg-de Vries type}, in {\it Current problems in
mathematics}, {\bf 24}, {\it Itogi nauki i tehniki},  VINITI, Moscow, 1984, 81-180, (in Russian), 
 {\it translation in J. Sov. Math.}, {\bf 30} (1985), 1975--2035.
%
\bibitem{Symm} P. J. Olver, {\it Applications of Lie Groups to Differential Equations,} Springer-Verlag, New York, 1989;
 N. H. Ibragimov, {\it  Applications of Transformation Groups to Mathematical Physics,}  Nauka, Moskow, 1983.
 
\bibitem{MSS} A. V. Mikhailov,  A. B. Shabat and  V. V. Sokolov, ``The symmetry approach  to classification of integrable equations'', 
{\it What is Integrability ?}, Springer-Verlag (Springer  Series  in Nonlinear Dynamics), New York, 1991, 115--184. 
%
\bibitem{SS}  V. V. Sokolov and S. I Svinolupov, {\it Math. Notes},  {\bf 48}:5-6 (1991),  1234--1239;
I.Sh. Akhatov, R.K. Gazizov,  N. H. Ibragimov, ``Nonlocal Symmetries. Heuristic Approach'',  {\it Itogi nauki i tehniki}, {\bf 34}, VINITI, Moscow, 
 1989, 3--83, (in Russian).
\bibitem{sergeev} A. Sergyeyev, ``On recursion operators and nonlocal symmetries of evolution equations'',
 {\it Hroc. Sem. Diff. Geom.}, Math. Publications V. 2, D. Krupka, ed., Silesian University in Opava, Opava, 2000, 159--173.
\end{thebibliography}
\end{document}